\documentstyle [12pt]{article}
\begin{document}
\title{The Emergence of Classicality via Decoherence: Beyond the
 Caldeira-Legget Environment
 \footnotetext{ to appear in Proceedings
 of the 4th Drexel Symposium on Quantum Nonintegrability}}
\author{Michael R. Gallis\\Department of Physics,
Penn State, Schuylkill Campus\\Schuylkill Haven, PA 17922 USA}
\date{}
\maketitle
\begin{abstract}
Maximally predictive states, as defined in recent work by
Zurek, Habib and Paz, are studied for more elaborate environment
models
than a linear coupling to an oscillator bath (which has become known
as
the Caldeira-Leggett model).  An environment model which includes
 spatial correlations in the noise is considered in the non-dissipative
regime.
 The Caldeira-Leggett model is also re-considered in the context of an
averaging
 procedure which produces a completely positive form for the quantum
master
equation.  In both cases, the maximally predictive states for the
harmonic
oscillator are the coherent states, which is the same result found by
Zurek,
Habib and Paz for the Caldeira-Legget environment.
 \end{abstract}

\section{Introduction}
Effective decoherence resulting from a quantum system's interaction
with
an environment can provide a natural mechanism for the transition from
quantum
 to classical behaviour for an open system.\cite{Joos&Zeh}
Decoherence has been an
integral part of several programs addressing the emergence of
classicality.\cite{Zurek,Halliwell}  One means of characterizing the
 effectiveness
of decoherence is the predictibility sieve, recently introduced by
Zurek, Habib
 and Paz (ZHP).\cite{ZHP}  They considered the particular model
consisting of
an independent oscillator bath linearly coupled to the system of
interest (often referred to as the
Caldeira-Leggett model\cite{CL}), for which they demonstrated that the
coherent states of
a harmonic oscillator are maximally predictive in that they correspond
to
minimal entropy  production under the effective dynamics of the open
system.

 I wish to consider the decoherence effects of more elaborate
environment
 models, motivated
by an extension of Joos and Zeh's scattering model\cite{Joos&Zeh}  to
arbitrary length
scales\cite{gallis&fleming1} and by the form of decoherence from
classical noise which
 includes spatial
correlations\cite{gallis1,diosi1}.  I will restrict discussion to the
Markov
regime, where the time correlation effects of the noise will not cause
memory
 effects in the effective quantum evolution.  Under these conditions,
the evolution can be written as
\begin{equation}
{ \partial \rho (x,x';t) \over \partial t}={\rm Hamiltonian}
+{\rm Dissipation}+\cdots-g(x,x')\rho (x,x';t).
\end{equation}
Decoherence is from the noise term, where
\begin{equation}
g(x;y) = {1\over{\hbar^2}}(c(x;x)+c(y;y)-2c(x;y)),
\end{equation}
is defined in terms of the random potential correlations:
\begin{equation}
\langle V({ x},t)V({ y},s)\rangle =c({ x};{y})\delta (t-s).
\end{equation}
The bulk of the models considered in the literature have
$g({ x};{y}) \propto (x-y)^2.$

In ZHP's scheme, maximally predicitive states are characterized by the
 production of linear entropy:$\, \varsigma  ( \rho  )={\rm Tr} [ \rho
 -{ \rho }^{ 2}]=1- Tr [{ \rho }^{ 2}].$ The rate of entropy
production
is then given by $\dot{\varsigma } ( \rho  )=- 2{\rm Tr}[ \rho
{\dot{\rho }}
 ]= - 2{\rm Tr}[ \rho { L} ( \rho  )]$ for the evolution operator $L$.
 Maximally predictive states are the pure states
$\rho  =|\mit \psi  )(\mit \psi |$
which have minimum entropy production
$\varsigma\rm =\int \dot{\varsigma}\rm \mit dt .$
The model system considered by ZHP was an oscillator linearly coupled
 to an oscillator bath (with factoring initial conditions).  For the
ohmic
 dissipation in the weak coupling, high temperature limit, the
effective
evolution is given by:
\begin{equation}
{\partial \rho  \over \partial t} = {1 \over i\hbar}[H,\rho
] -{{i \gamma}\over{2 \hbar}}[\{p,x\},\rho]
-{{2 m \gamma {k_B} T}\over{\hbar^2}}[x,[x,\rho]]
-{{i \gamma}\over \hbar}([x,\rho p]-[p,\rho x]).
\end{equation}
For weak dissipation, the second and fourth terms are ignored.  The
 first term corresponds to the usual unitary evolution, and the third
term to the noise which produces the decoherence.  In this case, the
instantaneous
rate of entropy production is:
\begin{equation}
 \dot{\varsigma} = {{8 m \gamma {k_B} T}\over{\hbar^2}}(<x^2>_{\psi}
-<x>_{\psi}^2) = 4D \Delta x ^{2},
\end{equation}
which when averaged over one full period $\tau$ of oscillation,
produces
an increase of entropy
\begin{equation}
\varsigma(\tau) = 2 D (\Delta x^2 + {{\Delta p^2}\over{m^2\omega^2}}).
\end{equation}
The minimization of the quantity in parentheses leads immediately
 to coherent states.
\section{Noise with Spatial Correlations}

Spatial correlation effects on decoherence can be considered most
easily
by looking at spatially correlated white noise using influence
functional
techniques.\cite{gallis1,diosi1}  The general characteristics of the
correlations in the context of decoherence via a scattering approach
have also been examined.\cite{gallis&fleming1}  Further motivation for
models
can be provided by examining nonlinear coupling to oscillator
baths,\cite{gallis3} such as for a particle locally coupled to an
 oscillator bath,
which has the lagrangian:
\begin{equation}
L =\int_{}^{}{ d}^{n}r \left\{{{1 \over 2}\left[{{\dot{ \phi }}
^{ 2}-{ c}^{ 2}({\nabla }_{ r}\phi  {)}^{2}}\right]+ \delta
  ({\bf r} -{\bf x} )\left[{{ m{\dot{\bf x}}^{2} \over 2}-
 \varepsilon \phi  ({\bf r}, t )- V ({\bf x} )}\right]}
\right\}
\end{equation}
The effective correlation function for the fluctuating
 forces (for $d$ dimensions)is given by
\begin{eqnarray}
 <{\bf  F}({\bf  x}, t  )\cdot {\bf  F}({\bf  y}, s )>&=&
<{\bf  F}({\bf r}={\bf  x}-{\bf  y}, \tau = t-s  )\cdot
 {\bf  F}  (0,0)> \nonumber \\
&=&{\hbar{ \varepsilon }^{ 2} \over
 2(2 \pi { )}^{ d}}\int_{ }^{}{ d}^{d}k {k}^{ 2}
\left\{{\coth({ \beta  \hbar \omega  \over  2})
 \over  \omega } \cos( \omega \tau )
\cos({\bf  k}\cdot({\bf  r}))\right\}
\end{eqnarray}
which can be contrasted with linear coupling where the forces
are perfectly correlated over all distances.

For a rather generic discussion, I consider a form of the correlation
which captures the relevant features of a wide variety of cases.  For
a one dimension case where the fluctuations and correlations are
 homogeneous (translationally invariant), isotropic and
fluctuations become uncorrelated at some characteristic length scale,
a reasonable model for the correlations is then
\begin{equation}
\langle V({ x},t)V({ y},s)\rangle ={\hbar}^{2}{\lambda \over 2}
{\rm e}^{-({{x-y}\over\sigma})^2} \delta (t-s).
\end{equation}
Additional dissipative effects will be ignored for this discussion.
The effective evolution is
\begin{equation}
{ \partial \rho (x,y;t) \over \partial t}={\rm Hamiltonian}
-g(x-y)\rho (x,y;t).
\end{equation}
The resulting ``decoherence rate'' is
\begin{equation}
 g(x-y) = \lambda (1-{\rm e}^{- ({{x-y}\over \sigma})^2}),
\end{equation}
which is quadratic at low length scales determined by the
parameter $\sigma$ (where linear models are relevant),
and which saturates to $\lambda$ for large length scales.
If decoherence is to be effective,
${\lambda \tau \geq 1}$ for dynamical timescales $\tau$.
For this model, the entropy production is
\begin{equation}
{\dot{\varsigma}} = -2 {\rm Tr}[\rho {\dot \rho}]
= -2 \int dx dy \rho(x,y) {\dot \rho}(y,x)
= -2 \int dx dy \rho(x,y) ({-g(y-x) \rho}(y,x)).
\end{equation}
For pure states $\rho(x,y) = \psi (x)
\psi^* (y)$ with $P(x) = \psi (x) \psi^* (x)$ and our choice of
$g(x;x')$
entropy production becomes
\begin{equation}
{\dot\varsigma} = 2 {\lambda \over {\hbar^2}}
\int dx dy P(x) P(y) {\rm e}^{- ({{x-y}\over \sigma})^2}.
\end{equation}
This expression does not lend itself to simple analysis, so I consider
two extreme cases.

For the first case, the environment correlation length scales
are taken to be
 much shorter than characteristic spread in $P(x)$.
In this case I find
\begin{equation}
{\dot \varsigma} \approx 2 {\lambda \over {\hbar^2}}
(1-\sigma {\sqrt{\pi}}\int  dx P^2 (x))
 \approx 2 {\lambda \over {\hbar^2}},
\end{equation}
that is, {\it all} states in this regime produce the same entropy, and
none are in any sense maximal.  Since ${\lambda \tau \geq 1}$, the
states
in this regime are rapidly being ``mixed'' by the noise, and have a
rapid
increase in entropy.  All states in this regime
have equally poor predictability.

For the second case, the environment correlation length scales
are taken to be
 much larger than characteristic spread in $P(x)$.  However, in this
case
the noise term is effectively quadratic, and the evolution is
approximately
\begin{equation}
{ \partial \rho (x,y;t) \over \partial t}={\rm Hamiltonian}
-{\lambda \over {\sigma \hbar^2}}(x-y)^{2} \rho (x,y;t),
\end{equation}
which is precisely the form used by ZPH in their analysis. In this
regime the coherent states are recaptured as the maximally
predicitive ones.

\section{Quantum Optical Master Equations from Bi-Linear Coupling to
Environment}

I would now like to consider alternate master equations which can be
obtained
for a harmonic oscillator with bi-linear coupling to the environment,
 using projection operator techniques in the weak
 coupling limit.\cite{alicki}
This is motivated in part by concerns regarding
 positivity of the master equation and
 intermediate temperatures,\cite{diosi2} as well as
the relevance of the details of transient behaviour and memory
 effects.\cite{gallis3}  Effective dynamics can be obtained by making
a naive assumption about the Markov nature of the dynamics, and
the evolution operator takes the form $ L = L_{o} + \Delta L$.
While an ohmic frequency distribution of environment oscillators taken
in
 the high temperature limit does reproduce the corresponding results
of Caldeira and Leggett,\cite{CL} this evolution is not positive.
However, an averaging procedure,
\begin{equation}
{\overline {\Delta L}}(\cdot)\equiv \lim_{a \to \infty}
{1 \over a} \int_{0}^{a} {\rm e}^{-L_o \tau} \Delta L
({\rm e}^{L_o \tau}(\cdot)) d\tau,
\end{equation}
does produce positive evolution, for arbitrary frequency
distributions
and for intermediate temperatures.\cite{alicki,gallisu,allan}
The resulting evolution turns out to be the Quantum Optical Master
Equation,
\cite{gardiner}
\begin{equation}
\Delta L = -{{\Gamma (\omega)}\over {4\hbar\omega}}
\big\{(2N+1)mw^{2}([x,[x,\rho]]
+{1\over{m^{2}\omega^{2}}}[p,[p,\rho]])
+2i\omega[x,\{p,\rho\}]+i\omega[\{p,x\},\rho]\big\},
\end{equation}
where
\begin{equation}
N = {1 \over {{\rm e}^{\beta\hbar\omega}-1}},
\end{equation}
and
\begin{equation}
\Gamma(w) = {\pi \over{2  \omega^{2} M}}
(n_{osc}{C^{2}\over m_{osc}})(\omega).
\end{equation}
In the low dissipation case, entropy production is given by
\begin{equation}
{\dot\varsigma}\propto \Delta x^{2} + {{\Delta
p^2}\over{m^{2}\omega^2},}
\end{equation}
so
\begin{equation}
\varsigma (\tau) \propto \Delta x^{2} + {{\Delta
p^2}\over{m^{2}\omega^2}.}
\end{equation}
 Again, coherent states are the maximal states when $\varsigma$
is minimized.  Thus the coherent states
are the ``most classical'' states for a variety of environments.

\end{document}